  \documentstyle[aps,epsfig]{revtex}

\newcommand{\mus}{\mbox{$\mu$s}}

\newcommand{\us}{\mbox{$\mu$s}}

\newcommand{\C}{\mbox{$^{12}$C}}

\newcommand{\N}{\mbox{$^{12}$N}}

\newcommand{\Ngs}{\mbox{$^{12}$N$_{\rm g.s.}$}}
\newcommand{\B}{\mbox{$^{12}$B}}

\newcommand{\Bgs}{\mbox{$^{12}$B$_{\rm g.s.}$}}

\newcommand{\numu}{\mbox{$\nu_{\mu}$}}
\newcommand{\numub}{\mbox{$\bar{\nu}_{\mu}$}}
\newcommand{\nue}{\mbox{$\nu_{e}$}}

\newcommand{\nueb}{\mbox{$\bar{\nu}_{e}$}}
\newcommand{\nutau}{\mbox{$\nu_{\tau}$}}

\newcommand{\nux}{\mbox{$\nu_{x}$}}
\newcommand{\nubx}{\mbox{$\bar{\nu}_{x}$}}

\newcommand{\ep}{\mbox{e$^{+}$}}

\newcommand{\el}{\mbox{e$^{-}$}}
\newcommand{\pos}{\mbox{e$^{+}$}}

\newcommand{\mum}{\mbox{$\mu^{-}$}}
\newcommand{\mup}{\mbox{$\mu^{+}$}}

\newcommand{\pip}{\mbox{$\pi^{+}$}}

\newcommand{\taum}{\mbox{$\tau^{-}$}}

\newcommand{\mupdecay}{\mbox{\mup\ $\rightarrow\:$ \pos $\!$ + \nue\ + \numub}}

\newcommand{\pipmup}{\mbox{\pip $\rightarrow\:$ \mup + \numu}}

\newcommand{\Ndecay}{\mbox{\Ngs\ $\rightarrow\:$ \C\ + \pos\ + \nue}}
\newcommand{\Bdecay}{\mbox{\Bgs\ $\rightarrow\:$ \C\ + \el\ + \nueb}}

\newcommand{\CB}{\mbox{\C\,(\,\nueb\,,\,\ep\,)\,\B }}
\newcommand{\CCprot}{\mbox{p\,(\,\nueb\,,\,\ep\,)\,n }}
\newcommand{\excl}{\mbox{\C\,(\,\nue\,,\,\el\,)\,\N$_{\rm g.s.}$}}

\newcommand{\NC}{\mbox{\C\,(\,$\nu$\,,\,$\nu'$\,)\,\C$^{*}$}}

\newcommand{\isov}{\mbox{\C\,(\,$\nu$\,,\,$\nu'$\,)\,\C$^{*}$
(\,1$^+$\,,\,1\,;\,15.1 MeV\,) }}

\newcommand{\etau}{\mbox{\nue $\leftrightarrow\,$\nutau }}
\newcommand{\emu}{\mbox{\nue $\leftrightarrow\,$\numu }}

\newcommand{\Aet}{\mbox{$A^{e\tau}$ }}
\newcommand{\Aem}{\mbox{$A^{e\mu}$ }}

\newcommand{\nuab}{\mbox{$\nu_{\alpha}\,\rightarrow\,\nu_{\beta}$ }}
\newcommand{\nuex}{\mbox{\nue $\rightarrow\,$\nux }}
\newcommand{\nuebx}{\mbox{\nueb $\rightarrow\,$\nubx }}
\newcommand{\nuenutau}{\mbox{\nue $\rightarrow\,$\nutau }}
\newcommand{\numunutau}{\mbox{\numu $\rightarrow\,$\nutau }}
\newcommand{\nuenumu}{\mbox{\nue $\rightarrow\,$\numu }}
\newcommand{\numux}{\mbox{\numu $\rightarrow\,$\nux }}
\newcommand{\numunue}{\mbox{\numu $\rightarrow\,$\nue }}
\newcommand{\numubnueb}{\mbox{\numub $\rightarrow\,$\nueb }}

\newcommand{\Pem}{\mbox{$P_{\nuenumu}$ }}

\newcommand{\Pex}{\mbox{$P_{\nuex}$ }}
\newcommand{\Ptau}{\mbox{$P_{\nuenutau}$ }}

\newcommand{\NCL}{\mbox{$90\%\,CL$}}

\newcommand{\Dm}{\mbox{$\Delta m^2$}}
\newcommand{\sit}{\mbox{$\sin ^2(2\Theta )$}}
\newcommand{\eVc}{\mbox{eV$^2$/c$^4$}}

\newcommand{\msig}[2]{\mbox{$\langle\,\sigma_{\rm #1}\,\rangle_{\rm #2}$}}
\newcommand{\msigma}{\mbox{$\langle\,\sigma\,\rangle$}}
\newcommand{\scc}{\mbox{$\sigma_{CC}\,(\,\nue\,)$}}

\newcommand{\snc}{\mbox{$\sigma_{NC}\,(\,\nue\,+\,\numub\,)$}}

\begin{document}

  \title{KARMEN Limits on \nuenutau\ Oscillations in 2--$\nu$ and 3--$\nu$ 
	 Mixing Schemes}

\author{B.~Armbruster$^a$, I.~Blair$^b$, B.A.~Bodmann$^c$, N.E.~Booth$^d$, 
G.~Drexlin$^a$, V.~Eberhard$^a$, J.A.~Edgington$^b$, C.~Eichner$^e$, 
K.~Eitel$^a$, E.~Finckh$^c$, H.~Gemmeke$^a$, J.~H\"o\ss l$^c$, T.~Jannakos$^a$, 
P.~J\"unger$^c$, M.~Kleifges$^a$,J.~Kleinfeller$^a$, W.~Kretschmer$^c$,
R.~Maschuw$^e$, C.~Oehler$^a$, P.~Plischke$^a$, J.~Rapp$^a$, C.~Ruf$^e$, 
M.~Steidl$^a$, O.~Stumm$^c$, J.~Wolf$^a$\cite{joa}, B.~Zeitnitz$^a$}

\address{$^a$ Institut f\"ur Kernphysik I, Forschungszentrum Karlsruhe,
  Institut f\"ur experimentelle Kernphysik, Universit\"at Karlsruhe,
  Postfach 3640, D-76021 Karlsruhe, Germany}
\address{$^b$ Physics Department, Queen Mary and Westfield College,
  Mile End Road, London E1 4NS, United Kingdom}
\address{$^c$ Physikalisches Institut, Universit\"at Erlangen-N\"urnberg,
  Erwin Rommel Strasse 1, D-91058 Erlangen, Germany}
\address{$^d$ Department of Physics, University of Oxford,
  Keble Road, Oxford OX1 3RH, United Kingdom}
\address{$^e$\it Institut f\"ur Strahlen- und Kernphysik, Universit\"at Bonn,
  Nu\ss allee 14-16, D-53115 Bonn, Germany}

\maketitle
  
\begin{abstract}
The 56\,tonne high resolution liquid scintillation calorimeter KARMEN at the
beam stop neutrino source ISIS has been used to search for neutrino oscillations
in the disappearance channel \nuex\ . The \nue\ emitted in \mup\ decay at rest
are detected with spectroscopic quality via the exclusive charged current 
reaction \excl\ almost free of background. Analysis of the spectral shape
of \el\ from the \nue --induced reaction as well as a measurement of the
absolute \nue\ flux allow to investigate oscillations of the type \nuenutau\
and \nuenumu . The flux independent ratio $R_{CC/NC}$ of charged current events 
\excl\ to neutral current events \NC\ provides additional information in
the oscillation channel \nuex . All three analysis methods show no evidence
for oscillations. For the \nuenutau\ channel \NCL\ limits of 
$\sit < 0.338$ for $\Dm \ge 100$\,\eVc\ and $\Dm < 0.77$\,\eVc\ for maximal 
mixing in a simple 2 flavor oscillation formalism are derived. A complete 3 
flavor analysis of the experimental data from five years of measurement
with respect to \etau\ and \emu\ mixing is presented.
  \end{abstract}


  \section{Introduction }\label{intro}
One of the most interesting issues in present particle physics is to 
clarify the neutrino mass problem. In the minimal standard model of electroweak
theory neutrinos are considered to be massless. But there is no compelling
theoretical reason behind this assumption. On the other hand, in most 
extensions of the standard model massive neutrinos are allowed. 
A very sensitive way of probing small 
neutrino masses and the mixing between different neutrino flavors is provided 
by neutrino oscillations. Although there are some experimental results 
pointing to neutrino oscillations, it is still difficult to deduce a 
consistent and reliable set of neutrino masses and 
mixings~\cite{acker,harris,fogli}. 

Up to now, experimental results have been persistently interpreted in terms of 
neutrino oscillations using only a 2--$\nu$ mixing scheme with a single 
mixing angle $\Theta$. However,
to perform a complete and consistent analysis of experimental data on
neutrino oscillation search, a 3 flavor formalism of neutrinos 
should be adopted. In this framework, all experiments
searching for different flavor oscillations (e.g. \numunue , \numubnueb ,
\numunutau\ in appearance mode or \nuex , \nuebx , \numux\ in disappearance 
mode) are combined to extract global information on the oscillation
parameters, the mixing angles as well as the mass differences
$\Dm_{ij} = |m^2_i - m^2_j|, i,j=1,\dots 3$. Searches for oscillations in the
disappearance mode, although less sensitive to small mixing angles, are of 
great importance to restrict the allowed parameter space.

In many of the 3 flavor descriptions of neutrino oscillations a so-called
'one-mass-scale dominance' $\delta m^2\equiv \Dm_{12}\ll \Dm_{13}, 
\Dm_{13}\approx \Dm_{23}\equiv \Dm$ has been adopted 
\cite{fogli,bilenky,babu,torrente,minakata,barshay,cardall}.
Possible mixing to sterile neutrinos as suggested by
\cite{suematsu,bereshiani,foot} is ignored whereas
CP conservation is assumed, as we shall do in the following. 
The flavor eigenstates $\nu_{\alpha}$ are then described by superpositions of
the mass eigenstates $\nu_i$ 
  \begin{equation}
    \nu_{\alpha} = \sum_{i=1}^3 U_{\alpha i}\nu_i \qquad (\alpha =e,\mu,\tau)
  \end{equation}
with the mixing matrix elements $U_{\alpha i}$ following the typical notation
\cite{pdg96} of the CKM-like mixing matrix described by two mixing angles 
$\Phi$,$\Psi$. The oscillation probability P$_{\alpha \beta}$ for 
transition \nuab\ in the KARMEN experiment with its 'short baseline' 
experimental configuration ($1/\Dm \approx L/E \ll 1/\delta m^2$) is then
  \begin{equation}
	P_{\alpha \beta} = A \cdot sin^2(1.27 \frac{\Delta m^2 L}{E})
  \label{osdef} \end{equation}
with \Dm\ in \eVc , the neutrino path length $L$ in meters and the neutrino
energy $E$ in MeV.
In a simple 2 flavor description $A=\sit$ with $\Theta$ being the mixing 
angle between the two $\nu$ states, while in a 3 flavor analysis
$A=4U_{\alpha 3}^2 U_{\beta 3}^2$ (appearance mode) or
$A=4U_{\alpha 3}^2 (1-U_{\alpha 3}^2)$ (disappearance mode) with
$U_{e3}^2=sin^2\Phi$, $U_{\mu 3}^2=cos^2\Phi sin^2\Psi$, 
$U_{\tau 3}^2=cos^2\Phi cos^2\Psi$ \cite{fogli}. 
For an experimental oscillation search 
it is important to note that the second factor in (\ref{osdef})
describing the spatial evolution of oscillations is unchanged in the 2 flavor
analysis. In the deduction of limits on $A$ and \Dm\ we shall start with
$A=\sit$ having in mind that the 
oscillation amplitude $A$ has different definitions depending on the actual 
theoretical model as well as the experimental situation (appearance
$\leftrightarrow$ disappearance search). In section~\ref{combi} and 
\ref{conclu} we shall then analyze our results in the 3 flavor scheme.

The {\bf KA}rlsruhe {\bf R}utherford {\bf M}edium {\bf E}nergy {\bf N}eutrino
experiment KARMEN has set stringent upper limits in the appearance channels 
\numubnueb\ ($\sit < 8.5\cdot 10^{-3}$ for $\Dm \ge 100$\,\eVc) and \numunue\
($\sit < 4.0\cdot 10^{-2}$ for $\Dm \ge 100$\,\eVc) \cite{karosci}.
KARMEN also searches for \nuenutau\ and \nuenumu\ through disappearance 
of \nue . Here, the neutrinos \nue\ emerge from \mup\ decay at rest (DAR)
and are detected at a mean distance of 17.7~m from the source. This search is
a complementary addition to the appearance channels. Compared to
other disappearance experiments this search has
very small, systematic uncertainties as well as an excellent 
signal to background ratio for the detection of \nue --induced events.
Limits on \nueb\ disappearance were obtained by reactor experiments e.g. at 
the nuclear power plant of Bugey~\cite{bugey} with \nueb\ from 
$\beta$-decays and the detector at 15, 40 or 95~m distance. Other limits on
\nuex\ were deduced from two $^{51}$Cr source experiments
in the GALLEX detector~\cite{gallex,bahcall} emitting \nue\ with 
discrete energies $E_{\nu}=0.746$\,MeV (81\%), 0.751\,MeV (9\%),
0.426\,MeV (9\%) and 0.431\,MeV (1\%). Experiments with neutrinos in the
energy range of several GeV allow the search for appearance \nue ,\numunutau .
In contrast to such experiments, like FNAL E531 \cite{ush86} with a 410 meter 
long decay path, in KARMEN the parameters $L$ and $E$ of $\nu$--induced events 
can be measured with high accuracy on an event-by-event basis.

The outline of the paper is as follows: 
We first present the experimental setup and the \nue\ detection and
identification (section \ref{K_setup}). 
Then we investigate the absolute number of identified 
CC reactions \excl\ (section~\ref{abscross}) comparing
the measured cross section with theoretical predictions. In a second 
evaluation method (section~\ref{crossratio}) we normalize the number of CC
sequences to the number of neutral current (NC) reactions \NC\,(15.1\,MeV).
These NC reactions are not sensitive to the $\nu$ flavor. 
By comparing the number of CC events with NC events 
measured simultaneously, systematic uncertainties in the calculation of
the absolute $\nu$ flux mostly cancel. A lower ratio of events 
R$_{exp}$=CC/NC than expected would also point to \nuex . 
A complementary analysis is given in
section~\ref{specshape}, where we search for distortions of 
the energy and spatial distributions of electrons from \excl , such as would
be due to \nuex . 
In section~\ref{combi} the results of these analyses are combined and
compared with other experimental limits for \nuebx , \nuenutau\ and 
\numunue (\numubnueb ) in a 2 flavor as well as a 3 flavor mixing scheme.

  \section{KARMEN Experiment}\label{K_setup}

  \subsection{Experimental Setup}
\begin{figure}
  \centerline{\epsfig{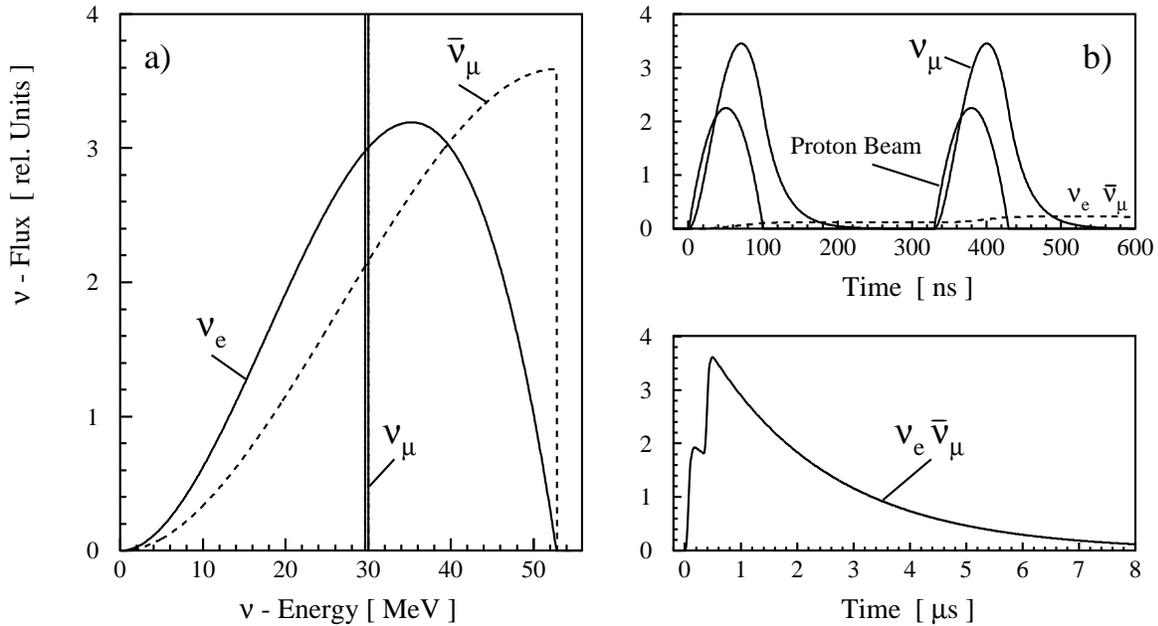}}
  \caption{Neutrino energy spectra (a) and production time (b) at ISIS.
	The proton double pulses are repeated with a frequency of 50\,Hz.}
  \label{isis_nu}
  \end{figure}

The KARMEN experiment is performed at the neutron spallation facility ISIS
of the Rutherford Appleton Laboratory. The rapid cycling synchrotron provides
800~MeV protons with an average current of 200\,$\mu$A. These are
stopped in a beam-stop Ta-D$_{2}$O-target producing neutrons and pions.
99\% of all charged pions are stopped within the target~\cite{bob}
leading to \pip\ decay at rest whereas negative pions are 
absorbed in the heavy target nuclei. Neutrinos therefore
emerge from the consecutive decay sequence \pipmup\ and \mupdecay\ in equal 
intensity per flavor. The intrinsic contamination \nueb/\nue\ from \mum\ decay
in the beam-stop-target is only $4.7\times 10^{-4}$~\cite{bob}.
The neutrino energy spectra are well defined due to the decay at rest
kinematics: The \numu\ from \pip --decay are monoenergetic 
($E_{\nu}$=29.8\,MeV), the continuous energy distributions of \nue\ and \numub\
up to $52.8$~MeV can be calculated using V--A theory (Fig.~\ref{isis_nu}a)
where the characteristic Michel shape of \nue\ from \mup\ DAR is given by the
expression~\cite{bou57}
  \begin{equation}
	\rm{N}(\epsilon)d\epsilon = 4\,\epsilon^{2}\,[\,3(1-\epsilon)\,+\,
	\frac{2}{3}\rho(4\epsilon-3)\,]d\epsilon \qquad
	\rm{with\;} \epsilon = \frac{\rm{E}_{\nu}}{\rm{E}_{\rm{max}}}
	\rm{\;and} \quad \rho = 0 \rm{\;for\;} \nue
  \end{equation}
The neutrino production time follows the two parabolic proton pulses of 
100\,ns base width and a separation of the maxima of 324\,ns, produced
with a repetition frequency of 50\,Hz. The different lifetimes of pions
($\tau$\,=\,26\,ns) and muons ($\tau$\,=\,2.2\,$\mu$s) 
together with this unique time structure allow a clear
separation in time of the \numu -burst from the following \nue\ and
\numub\ (Fig.~\ref{isis_nu}b) which show the characteristic exponential
distribution of the muon lifetime $\tau$\,=\,2.2\,$\mu$s. 
Furthermore the accelerator's extremely small duty cycle suppresses
effectively any beam-uncorrelated background by four to five 
orders of magnitude depending on the time interval selected for analysis.

The neutrinos are detected in a high resolution 56\,t liquid scintillation
calorimeter segmented into 512 central modules with a cross section of 
$0.18\times 0.18$\,m$^2$ and a length of 3.53\,m each~\cite{karmen_det}.
A massive blockhouse of 7000\,t of steel in 
combination with a system of two layers of active veto counters provides 
shielding against beam correlated spallation neutron background and suppression
of the hadronic component of cosmic radiation, as well as highly efficient
identification of cosmic muons and their interactions.
The central scintillation calorimeter and the inner veto counters are
segmented by double acrylic walls with an air gap providing efficient light
transport via total internal reflection of the scintillation light at the
module walls. The event position within an individual module is determined 
with a position resolution of $\pm$\,6\,cm for typical energies of 30\,MeV 
\cite{dodd} by
the time difference of the PM signals at each end of this module. Due to the 
optimized optical properties of the organic liquid scintillator \cite{scinti} 
and an active volume of 96\% for the calorimeter, an energy resolution of
$\sigma_E=11.5\%/\sqrt{E [MeV]}$ is achieved.
The KARMEN electronics is synchronized to the ISIS proton pulses to an
accuracy of better than $\pm 2$\,ns, so that the time structure of the
neutrinos can be exploited in full detail.

  \subsection{Detection and Identif\/ication of \nue }

Electron neutrinos \nue\ from \mup\ decay at rest (DAR) are detected via the 
exclusive charged current (CC) reaction \excl\ which leads to a 
delayed coincidence signature. The ground state of \N\ decays at its production
point with a lifetime of $\tau = 15.9$\,ms: \Ndecay . The detection signature
consists of a prompt electron within a few \mus\ after beam-on-target
($0.6\le t_{pr}\le 9.6$\,\us) followed by a spatially correlated 
positron (\el/\ep\ sequence with $0.5\le t_{\rm diff}\le 36$\,ms). 
To reduce cosmic background, no event is accepted within a software deadtime
of 20\,\us\ after any signal in the detector system. Remaining sequences 
induced by cosmic background are studied with high statistical accuracy
by demanding the prompt event to emerge in a 200\,\us\ long time interval 
before beam--on--target when no neutrinos are produced at ISIS. This CC 
reaction has been previously studied in detail \cite{karmen_cc1}, with respect 
to weak nuclear form factors \cite{karmen_cc2} and to the CC helicity 
structure of muon decay \cite{karmen_om}. 

\begin{figure}
  \centerline{\epsfig{figure=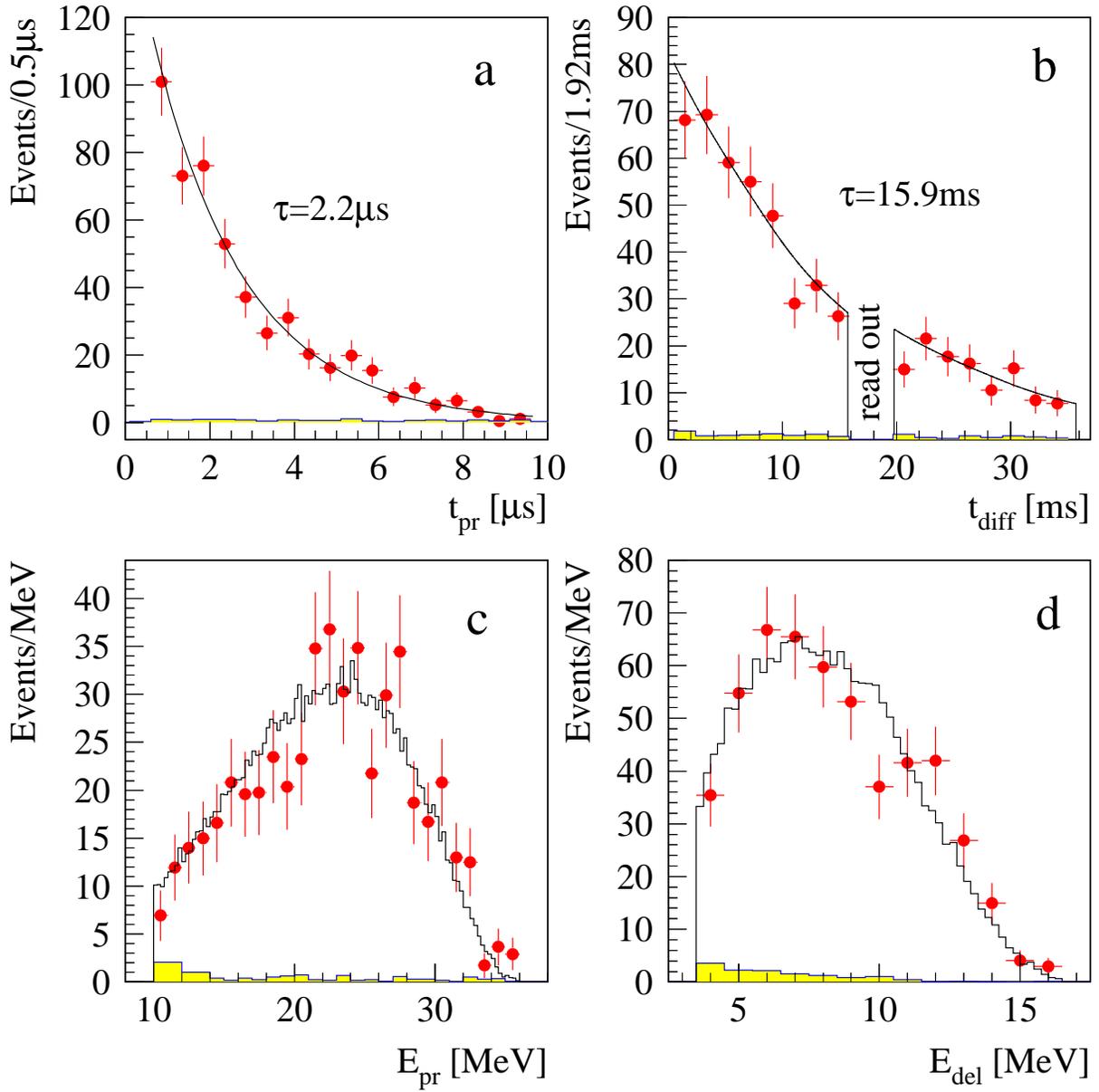,width=16.0cm}}
  \caption{Signatures of the \nue\ detection reaction \excl :
	time and energy distributions of the prompt electron (a;c) 
	and the delayed positron (b;d).
	Data points are with background subtracted, MC expectations are shown
	as histograms, total background as shaded histograms.}
  \label{CC_all}
  \end{figure}

The data set used for this \nuex\ analysis is taken from 1990--1995
corresponding to 9122\,C protons on the ISIS beam stop target, or 
$2.51\times 10^{21}$  \mup\ decays at rest in the target.
Figure~\ref{CC_all} shows the signatures of 513 events surviving all cuts.
These events contain $499.7\pm 22.7$ $\nu$--induced \el/\ep\ sequences and 
$13.3\pm 0.8$ cosmic induced background sequences.
The energy spectra of the prompt \el\ and the delayed \ep\ follow the Monte 
Carlo (MC) expectations from \mup\ DAR and the \N\ $\beta$--decay.
The time distribution of the prompt event reflects the muon lifetime,
the time difference $t_{\rm diff}$ shows the \N\ lifetime folded with the
detection efficiency. 

The stringent signatures of the \excl\ reaction lead to an excellent
signal--to--background ratio of 37:1 which allows identification of \nue\ 
almost free of background. Accordingly, the search for \nue\ disappearance
has very small systematic errors and the sensitivity in this oscillation
channel is not limited by background events.

  \section{Oscillation Searches }\label{oscisearch}

Flavor oscillations of the type \nuenumu\ or \nuenutau\ will result in a
smaller number of detected \excl\ sequences than expected because
the analogous charged current reactions 
\mbox{\C\,(\,\numu\,,\,\mum\,)\,\N$_{\rm g.s.}$}
and \mbox{\C\,(\,\nutau\,,\,\taum\,)\,\N$_{\rm g.s.}$} cannot be induced due
to the small neutrino energies ($E_{\nu}\le 52.8$\,MeV). 
In addition to the comparison of the observed \nue\ rate with expectation, 
the effects of \nue\ disappearance can be investigated by a detailed study
of the remaining \nue --induced events. First,
the characteristic energy dependence of flavor oscillations
(\ref{osdef}) will lead to a distortion of the electron energy distribution.
Secondly, the $1/r^2$ dependence of the \nue\ flux from \mup\ DAR in the beam 
stop target will be distorted. The specific features of 
neutrino production at ISIS and detection by KARMEN require the two
oscillation channels \nuenumu\ and \nuenutau\ to be discussed separately.

We first consider transitions \nuenutau . As described above, \nutau\ cannot 
be detected via CC reactions, so oscillations \nuenutau\ directly lead to a 
reduction of CC sequences and consequently of the measured flux averaged 
cross section \msig{CC}{exp}.

\begin{figure}
  \centerline{\epsfig{figure=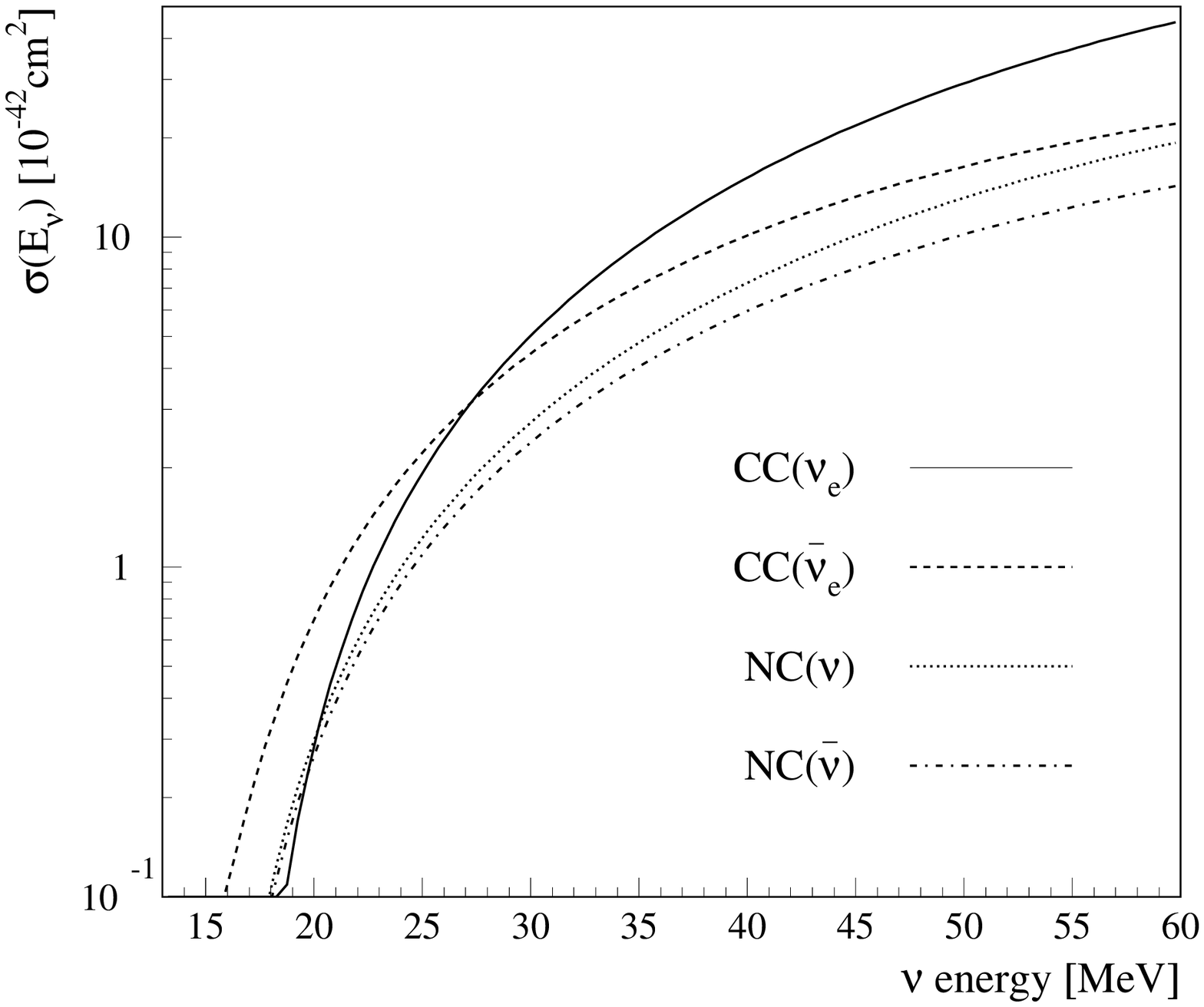,width=16.0cm}}
  \caption{Energy dependence of the cross sections of charged current (CC) 
	reactions \excl\ and \CB\ and of neutral current (NC) \isov\ for 
	different $\nu$ flavors in an elementary particle treatment of the 
        Carbon nucleus \protect\cite{fuku} . }
  \label{CC_cross}
  \end{figure}

\begin{figure}
  \centerline{\epsfig{figure=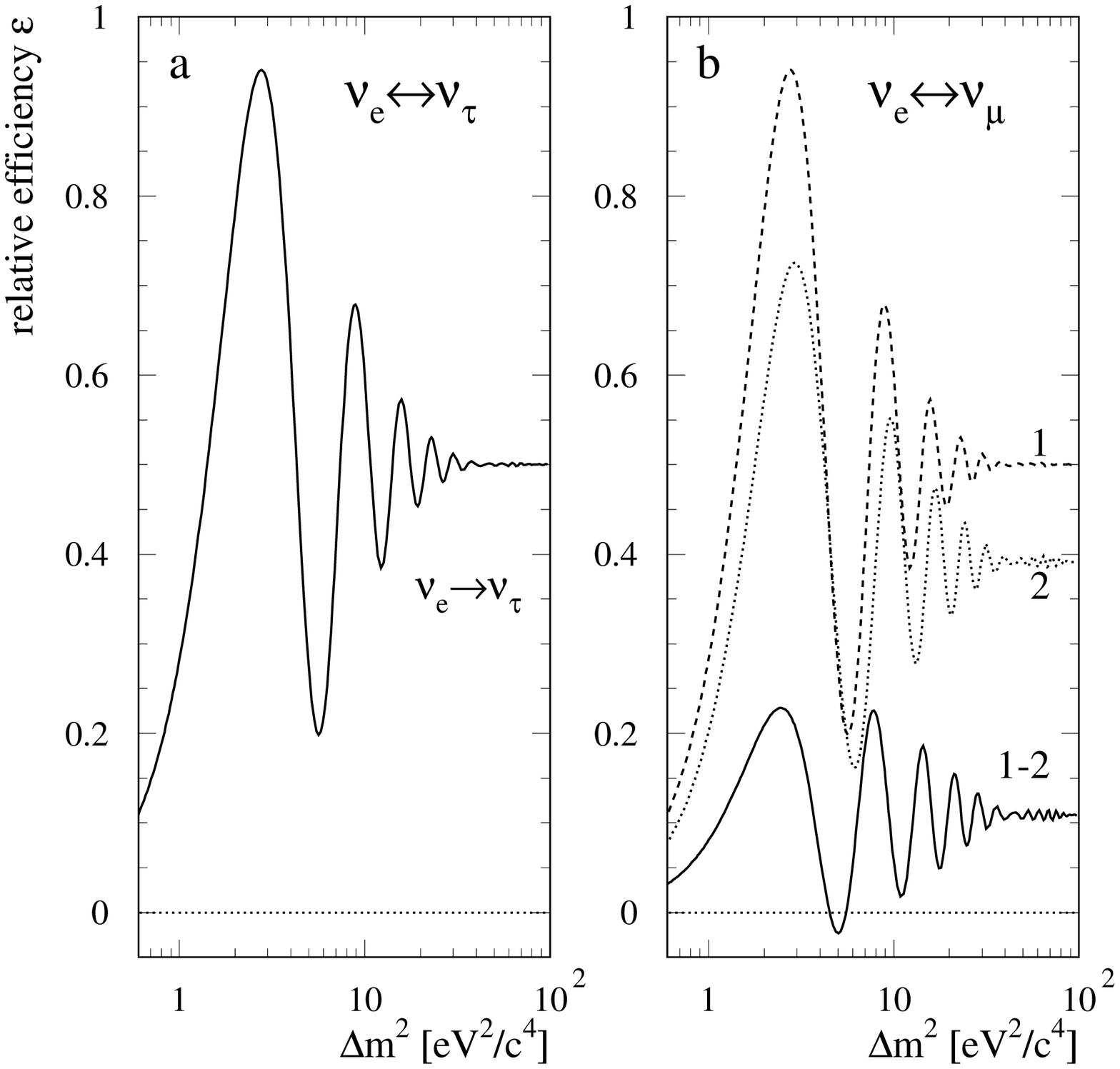,width=16.0cm}}
  \caption{Relative detection efficiencies $\epsilon$ as functions of the mass
	parameter \Dm : a) $\epsilon(\nu)$ for \etau\ mixing; b) \emu\ mixing:
	curve 1 shows $\epsilon(\nu)$ for \nuenumu\ disappearance, curve 2
	shows $\epsilon(\bar{\nu})$ for \numubnueb\ appearance; curve 1--2 
	shows the net efficiency $\epsilon(\nu)-\epsilon(\bar{\nu})$ for
	the combination of \nuenumu\ and \numubnueb . The efficiencies 
	$\epsilon$ are normalized to the detection efficiency of \excl\ 
	sequences within the applied cuts.}
  \label{CC_eff}
  \end{figure}

The second channel, \emu\ mixing, is more complex due to the simultaneous
presence of \numub\ in the neutrino beam from \mup\ decay. Assuming CPT 
invariance, there would be a second source for 
CC reactions in the KARMEN detector, namely \nueb\ from \numubnueb\ leading 
to \CB\ since \numub\ from \mup\ DAR at ISIS are 
produced with the same intensity and at the same time as \nue .
These would mostly compensate the reduction through \nuenumu\ of the measured 
cross section due to the following reasons:

Because \B\ belongs to the isobar triplet \B -\C -\N , 
the Q--value for \CB\ ($Q=-13.4$\,MeV) as well as the \B\ lifetime
($\tau=29.14$\,ms) and the endpoint energy of \Bdecay\ ($E_0=13.4$\,MeV)
are very similar to the corresponding quantities for \excl\ 
with the subsequent \N\ $\beta$-decay (see Fig.~\ref{CC_n12b12}).
Moreover, the cross sections are similar with 
$\msigma\ = 9.0\cdot 10^{-42}$\,cm$^2$ \cite{fuku} induced by \nueb\ from 
\numubnueb\ with large \Dm\ (see Fig.~\ref{CC_cross}).
So, as one CC reaction disappears, the other takes its place. 
\footnote{The detection of \numubnueb\ oscillations by \CCprot\ leads to a
different signature as the neutron capture time $\tau_n=120$\,\us\ is much 
faster than the $\beta$--decays of \N\ or \B\ (0.5--36\,ms time interval).}
Figure~\ref{CC_eff} demonstrates this argument in detail showing the
relative detection efficiencies $\epsilon(\nu)$ and $\epsilon(\bar{\nu})$
as a function of the oscillation parameter \Dm . Taking into account
the different energy dependent cross sections for both CC reactions, the
detection efficiency $\epsilon(\nu,\Dm)$ within the KARMEN detector for \nuex\ 
disappearance into \nutau\ or \numu\ is only slightly higher than for 
\numubnueb\ appearance $\epsilon(\bar{\nu},\Dm)$ which also incorporates 
the different detection efficiency for the sequential \ep/\el\ due to the 
different lifetimes and Q-values of \N/\B .
The different $\nu$ energy spectra and energy dependences of the cross sections 
are responsible for a small shift in the position of minima and maxima of the 
efficiencies for \nuex\ and \numubnueb . Therefore, at $\Dm \approx 5$\,\eVc\ 
a small net increase of CC reactions is expected assuming
mixing between \nue\ and \numu .

Following these considerations the performed search for \nuex\ must actually be 
interpreted in terms of \emu\ and \etau\ mixing where the 
\nuenutau\ search is much more sensitive ( in terms of the absolute number 
of CC events; see section \ref{abscross}) than \nuenumu\ due to the almost 
complete cancellation by the mirror \numubnueb . 
An analysis of the spectral shape of CC sequences, 
however, is sensitive to both mixings, \etau\ and \emu , as will be shown in 
section~\ref{specshape}. This shape analysis is performed in a complete 3 
flavor scheme where the amplitudes for the mixings \etau\ and \emu\ are 
completely free and independent parameters.

  \subsection{Absolute Cross Section }\label{abscross}

\begin{figure}
  \centerline{\epsfig{figure=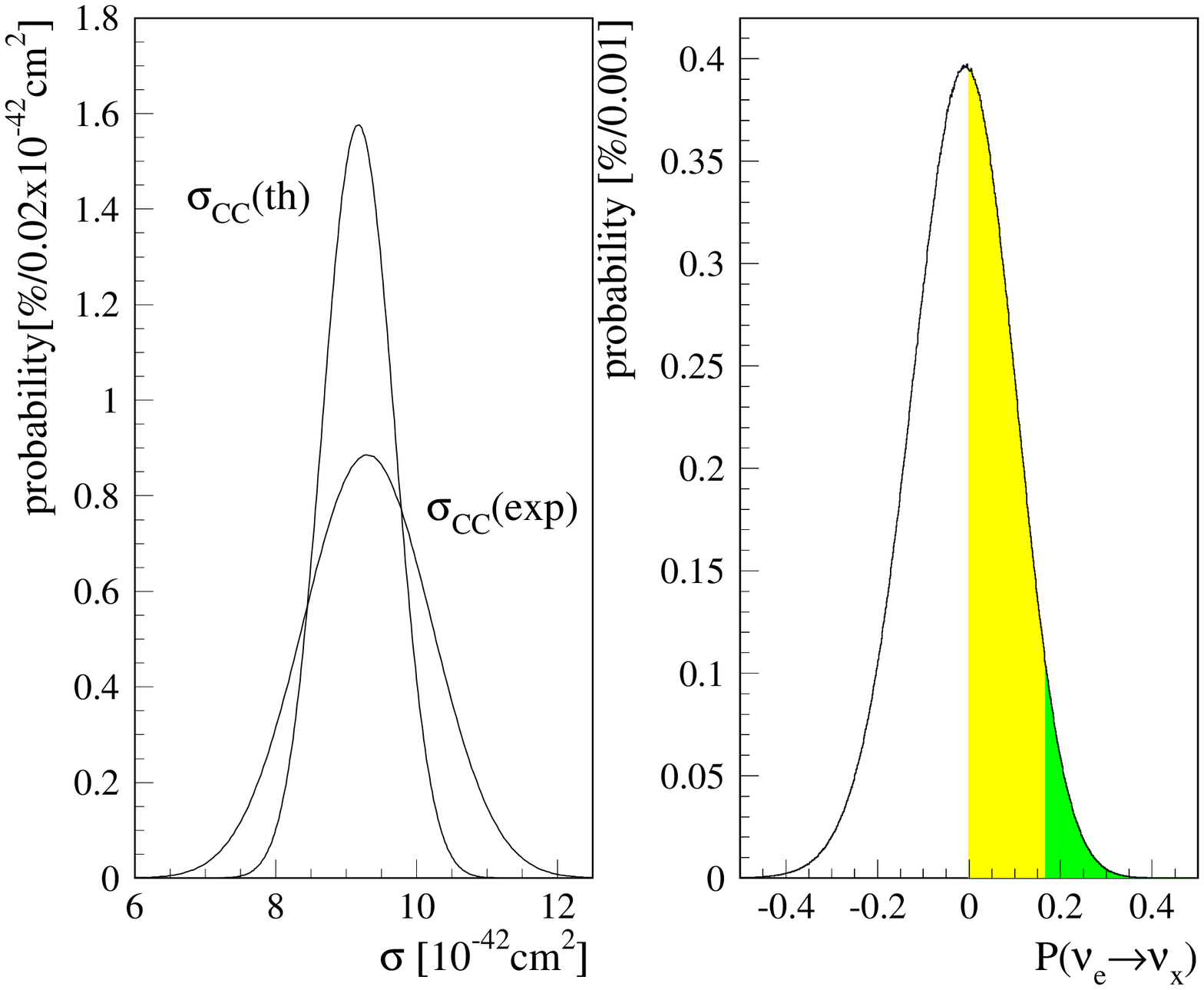,width=14.0cm}}
  \caption{Simulated distributions of cross sections \msig{CC}{exp} and
	\msig{CC}{th} of \excl\ and resulting \Pex\ according
	to Gaussian distributed values of \msig{CC}. The shaded areas show the
	physical region, darkly shaded are the upmost 10\% of the shaded
	\Pex\ distribution.}
  \label{P_sim}
  \end{figure}

The comparison of the number of \nue --induced CC reactions with the expectation
from theoretical calculations is done on the basis of the measured flux
averaged cross section \msig{CC}{exp} for \nue\ from \mup\ DAR. 
Taking into account the overall detection efficiency of $\epsilon_{CC}=0.328$
the 499.7$\pm$22.7 \nue --induced events (see Fig.~\ref{CC_all}) correspond to 
a flux averaged cross section of
\begin{equation} 
   	\msig{CC}{exp} = ( 9.4 \pm 0.4_{(stat.)} \pm 0.8_{(syst.)} ) 
	\times 10^{-42} {\rm cm}^2 
\end{equation}
where the systematic error is almost entirely due to the uncertainty 
in the calculation of the absolute $\nu$ flux from ISIS~\cite{bob}. 
The measured cross section is in excellent agreement with theoretical 
calculations based on different models to describe the \C\ nucleus (see 
table~\ref{crosstable}), i.e. there is no indication of \nue\ disappearance.
For quantitative calculations we use the mean theoretical value
of \msig{CC}{th} from \cite{fuku,mintz,kolbe94,kolbe96,conn}
with a realistic estimate of the systematic error~\cite{vogel}: 
$\msig{CC}{th} = ( 9.2 \pm 0.5 ) \times 10^{-42}$\,cm$^2$.
The \nue\ disappearance oscillation probability \Pex\ can be written
in terms of cross sections as
\begin{eqnarray}
\Pex & = & 1 - \frac{\msig{CC}{exp}}{\msig{CC}{th}} \\ 
     & = & 1 - \frac{9.4 \pm 0.89}{9.2 \pm 0.5}  \label{ratio} \end{eqnarray}
where the statistical and systematical errors of the experimental value have
been added quadratically. To get a density distribution for
values of \Pex\ we sampled \msig{CC}{exp} and \msig{CC}{th} from Gaussian
distributions with the given mean values and widths (Fig.~\ref{P_sim}). 
The resulting distribution of \Pex\ is slightly non Gaussian due to the ratio 
introduced in (\ref{ratio}). Only positive values of \Pex\ correspond to 
the physically allowed region of oscillation. To extract a \NCL\ upper limit we 
therefore renormalize the distribution in \Pex\ following the most conservative
Bayesian approach \cite{pdg96}. The limit deduced is then
\begin{equation} 
	\Pex < 0.169 \qquad ( \NCL ) \label{CC_abs_result} 
\end{equation}
The interpretation of this limit in terms of the oscillation parameters 
\Dm\ and mixing angle is given in section \ref{combi}.

\subsection{Ratio of Cross Sections CC/NC }\label{crossratio}

KARMEN measures, simultaneously, the CC reaction \excl\ with its 
sequential \el/\ep\ signature and
the NC reaction \isov\ induced by \nue ,\,\numub\ from \mup\ DAR.
The NC event signature consists of a $15.1$\,MeV $\gamma$ from deexcitation
of \C $^*$ detected within the \nue ,\,\numub\ time window after beam-on-target.
This reaction was observed by KARMEN for the first time \cite{karmen_nc}
and is described in detail in \cite{karmen_nu,erice97}. 
The updated cross section for the 1990--1995 data
\begin{equation} 
   	\msig{NC}{exp} = ( 10.8 \pm 0.9_{(stat.)} \pm 0.8_{(syst.)} ) 
	\times 10^{-42} {\rm cm}^2 
\end{equation}
again is in very good agreement with theoretical calculations 
(table~\ref{crosstable}).

In the case of oscillations between \nue\ and other $\nu$ flavors the NC cross
section remains unchanged due to the flavor universality of NC interactions.
\footnote{Note that this argument is only valid for flavor oscillations 
with neutrinos \nue ,\numu ,\nutau . Neutrino oscillations 
\mbox{\nue $\rightarrow\,\nu_s$} involving sterile neutrinos would reduce
both the number of \nue --induced CC as well as \nue ,\numub --induced NC
reactions.}
The number of CC events \excl\ would decrease due to \nuenutau . By looking at
the ratio of observed events (i.e. the measured cross sections) the
systematic uncertainties in \msig{}{exp} are significantly reduced so that
statistical fluctuations dominate the error on $R_{exp}$:
\begin{equation} 
	R_{exp} = \frac{\msig{CC}{exp}}{\msig{NC}{exp}} = 0.86 \pm 0.08 
\end{equation}
This $\nu$ flux independent ratio $R_{exp}$ is in good agreement with
theoretical predictions (see table~\ref{crosstable}). Again, there is no
indication of \nue\ disappearance.

As theoretical prediction for $R_{th}$ we combine the $R$--values from 
\cite{fuku,mintz,kolbe94,kolbe96,conn} in a straight forward way
and take the mean value and its variance as an estimate for the systematic 
error: $R_{th} = 0.91 \pm 0.035$. The \nuex\ oscillation probability 
can then be written as 
\begin{eqnarray}
	\Pex & = & 1 - R_{exp}/R_{th} \\
	      & = & 1 - \frac{0.86 \pm 0.08}{0.91 \pm 0.035}
\end{eqnarray}
In close analogy to the evaluation in section~\ref{abscross} an upper limit 
can be extracted after renormalization to the physical region:
\begin{equation} 
	\Pex < 0.192 \qquad ( \NCL ) \label{CC_NC_result} 
\end{equation}
This limit is slightly higher than the limit (\ref{CC_abs_result}) in section 
\ref{abscross} due to the relatively high value of the experimental cross 
section \msig{NC}{exp} compared to the theoretical predictions.

\subsection{Spectral Shapes }\label{specshape}

\begin{figure}
  \centerline{\epsfig{figure=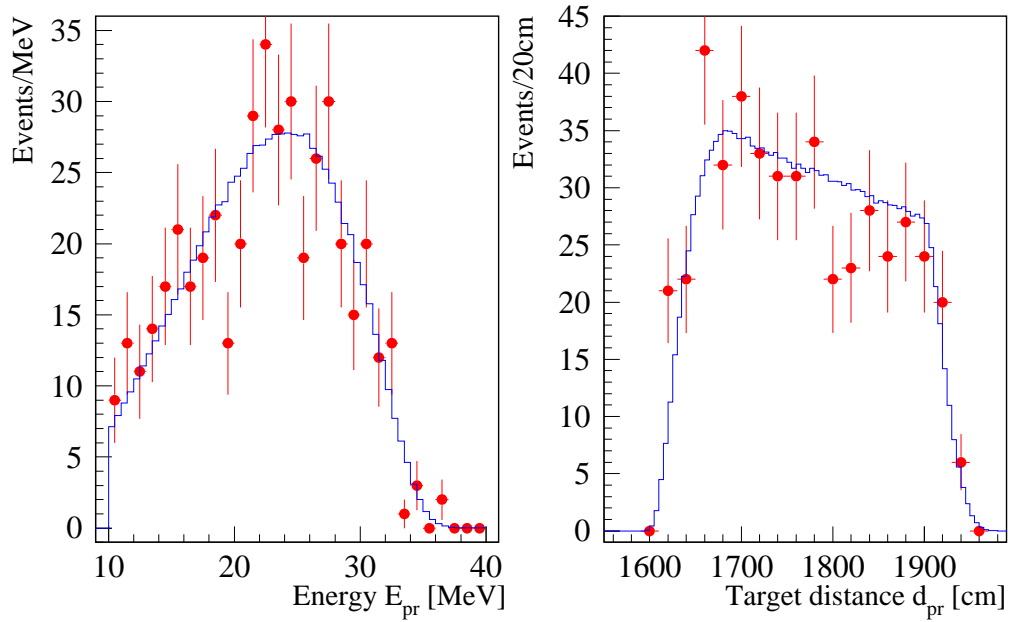,width=14.0cm}}
  \caption{Distributions of energy and target distance of prompt events 
	(data points) with stringent cuts for the \nuex\ shape analysis and 
	underlying MC expectations without oscillations.}
  \label{evts_le}
  \end{figure}

The third means of searching for oscillations \nuex\ is to carry out a
maximum likelihood (ML) analysis of
the measured energy and spatial distribution of the prompt electrons from
\excl\ reactions. 
This method is not sensitive to the absolute number of detected
sequences but to distortions of the expected spectra in energy and position 
within the detector volume. 
In order to minimize contributions from cosmic induced background we
applied more stringent cuts in time (corresponding to 3\,$\tau_{\mu^+}$) and 
spatial distribution (restricted fiducial volume of 86\% of the central
detector) for the prompt event resulting in 458 accepted sequences
(see Fig.~\ref{evts_le}) with 5.8 background events determined in the 
appropriate prebeam evaluation. 
With a neutrino signal--to--background ratio of about 80:1 the cosmic background
can be included as a fixed component in the likelihood analysis and will not
be discussed further. The aim of the ML analysis is
to calculate a combined likelihood for all 458 events on the basis that
the two--dimensional density distribution $f(L,E_{\el})$ is sensitive to 
varying amounts of oscillation events from \nuenutau , \nuenumu\ and 
\numubnueb . $E_{\el}$ is related to neutrino energy by
$E_{\el}=E_{\nue}-17.3$\,MeV, $L_{\el}=L_{\nue}$ since the electron is
detected at the interaction point. The absolute energy scale is known to a
precision of $\pm 0.25$\,MeV from analysis of Michel electrons from decay of
stopped cosmic muons. The density function $f$ for no oscillations is 
  \begin{equation} 
	f_{no osc.}(L,E_{\el}) = \frac{1}{F_{no osc.}} \cdot
        n(L) \cdot \Phi_{\nue}(E_{\nue}) \cdot \sigma(E_{\nue})
  \label{stan_dens_theo} \end{equation}
with the normalization factor $F$ integrated over the \nue\ energy and the
detector volume
  \begin{equation} 
	F_{no osc.} = \int_{E_{\nue}} \int_{Det}
        n(L) \cdot \Phi_{\nue}(E_{\nue}) \cdot \sigma(E_{\nue}) dL dE_{\nue}
  \label{int_dens_theo} \end{equation}
The function $n(L)$ represents the spatial distribution of events within
the rectangular detection volume resulting from an isotropic neutrino flux 
from ISIS ($\Phi_{\nu}\sim L^{-2}$). $\Phi_{\nue}$ denotes the \nue\ energy
distribution from \mup\ DAR (see Fig.~\ref{isis_nu}a),
$\sigma(E_{\nue})$ the energy dependent cross section for \excl\ 
(see Fig.~\ref{CC_cross}). 

\begin{figure}
  \centerline{\epsfig{figure=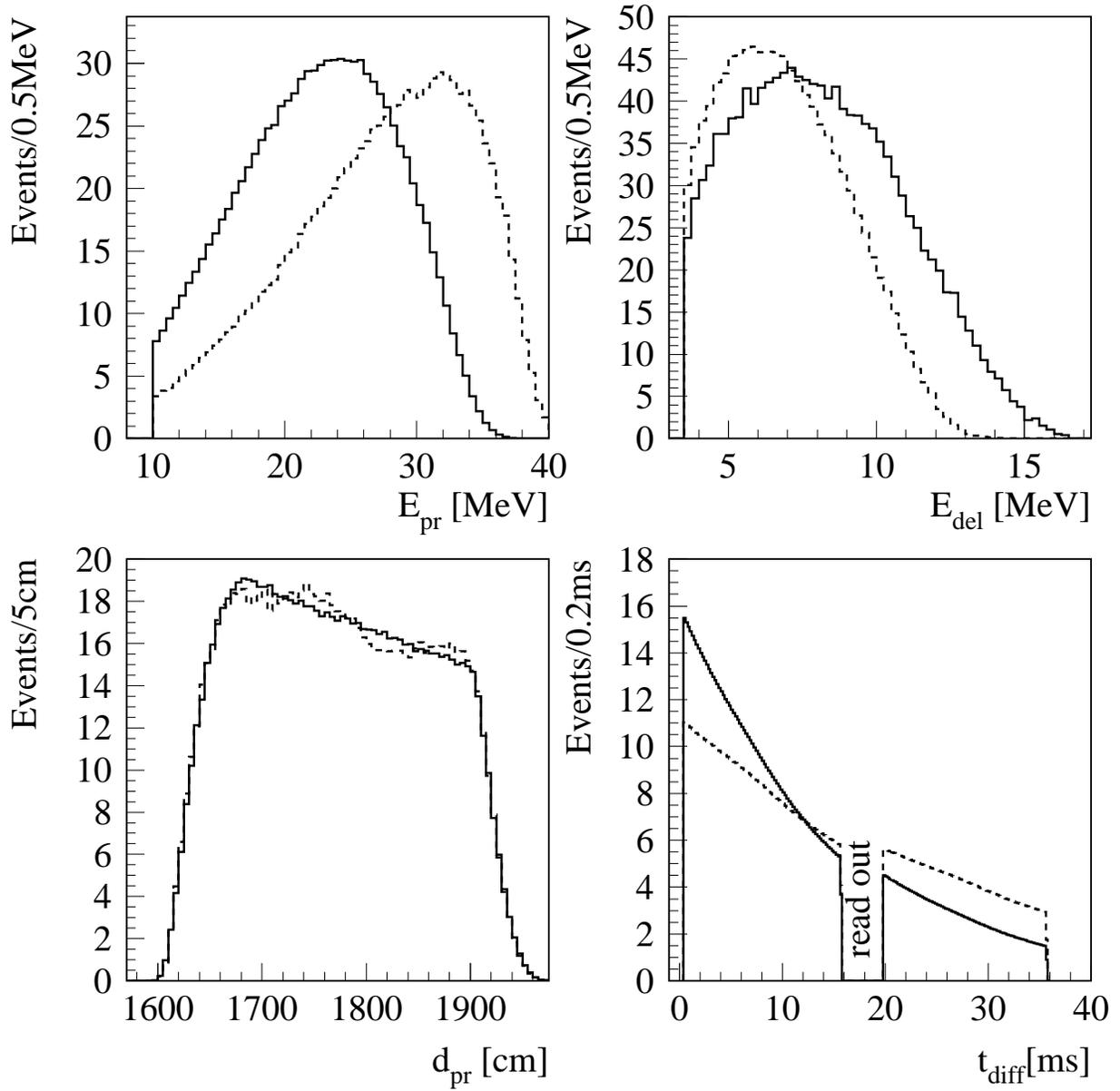,width=16.0cm}}
  \caption{Simulated detector response in energy, target distance 
	and time difference including readout efficiency of \excl\ (solid 
	histo\-grams) and \CB\ (dashed histograms) for \Dm\,=\,100\,\eVc\ 
	normalized to the same arbitrary number of events.}
  \label{CC_n12b12}
  \end{figure}

\begin{figure}
  \centerline{\epsfig{figure=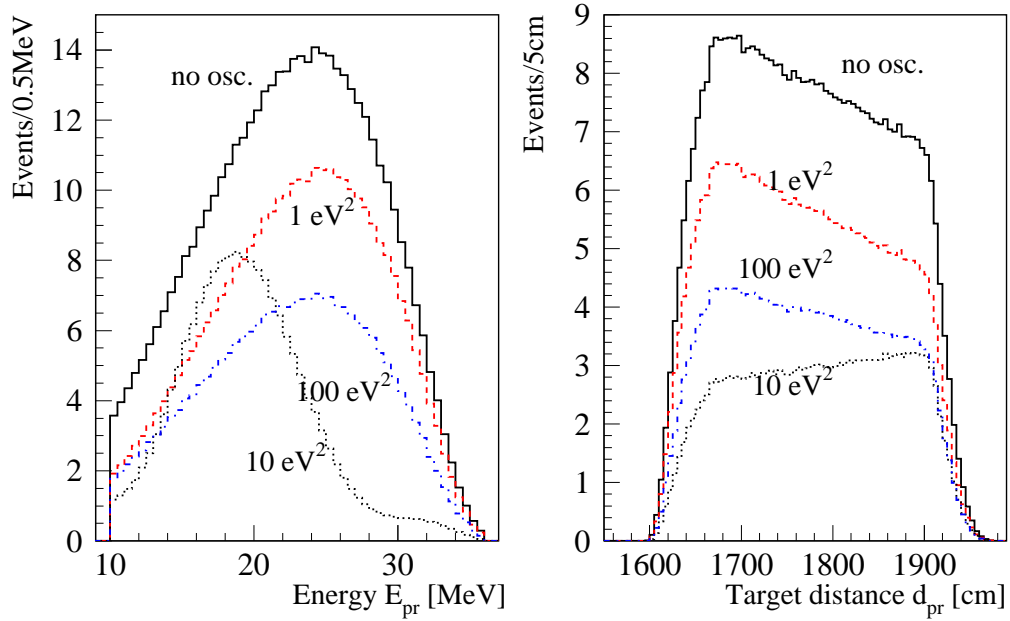,width=14.0cm}}
  \caption{Examples of simulated \el\ energy and distance distributions
	($E$-- and $L$--projection of $f_{noosc.}-f^{\nu}_{\Delta m^2}$) in
	the oscillation scenario \nuenutau\ for maximal mixing $\Aet =1$
	with different
	\Dm\ as well as for no oscillation. The spectra show the 
	visible electron signal in the detector and are normalized to the
	experimental event number $N=458$ for no oscillation 
	(see Fig.~\ref{evts_le}). }
  \label{MC_le}
  \end{figure}

For an oscillation \nuex\ with given \Dm\ the
density function $f^{\nu}_{\Delta m^2}$ normalized by the integral 
$F^{\nu}_{\Delta m^2}$ over $L$ and $E$ can be written as
  \begin{equation} 
	f^{\nu}_{\Delta m^2}(L,E_{\el}) = \frac{1}{F^{\nu}_{\Delta m^2}} \cdot
	sin^2(1.27 \frac{\Dm L}{E_{\nue}})
        \cdot n(L) \cdot \Phi_{\nue}(E_{\nue}) \cdot \sigma(E_{\nue})
  \label{osci_dens_theo} \end{equation}
The appearance of \nueb\ through \numubnueb\ is described by the density
  \begin{equation} 
	f^{\bar{\nu}}_{\Delta m^2}(L,E_{\ep}) = 
	\frac{1}{F^{\bar{\nu}}_{\Delta m^2}} \cdot sin^2(1.27 
	\frac{\Dm L}{E_{\numub}})
	\cdot n(L) \cdot \Phi_{\numub}(E_{\numub}) \cdot \sigma(E_{\nueb})
  \label{oscib_dens_theo} \end{equation}
in close analogy to (\ref{osci_dens_theo}).
The density functions used in the ML analysis are MC--simulated
electron/positron (see Fig.~\ref{MC_le}) and positron/electron 
(see Fig.~\ref{CC_n12b12}) distributions including all effects due to detector 
resolutions and the cuts applied .
The combined likelihood for the $N=458$ events is calculated with respect to 
the fraction $r_1$ of initially expected electrons now absent due to
\nuenutau\ and $r_2$ of the combined transition \nuenumu\ and \numubnueb\
taking into account the different detection efficiencies $\epsilon(\nu,\Dm)$ and
$\epsilon(\bar{\nu},\Dm)$ (see Fig.~\ref{CC_eff}):
  \begin{eqnarray}
	L_{\Delta m^2}(r_1,r_2) &=& \prod_{k=1}^{N} L_{\Delta m^2}^k(r_1,r_2)
 	= \prod_{k=1}^{N} \: \{ \: (1+r_1+r_2)\cdot f_{noosc.}(E_k,L_k) 
	\nonumber \\ &-& r_1\cdot f^{\nu}_{\Delta m^2}(E_k,L_k)
	- r_2\cdot \frac{1}{\epsilon(\nu,\Delta m^2) - 
	\epsilon(\bar{\nu},\Delta m^2)} \nonumber \\ & \cdot &
	[ \: \epsilon(\nu,\Delta m^2) \cdot f^{\nu}_{\Delta m^2}(E_k,L_k) - 
	\epsilon(\bar{\nu},\Delta m^2) \cdot f^{\bar{\nu}}_{\Delta m^2}(E_k,L_k) 
	\: ] \: \}
  \label{lhd_sum} \end{eqnarray}
In this specific construction of the likelihood function $L$ the event number
is fixed at $N=458$, only the shape differences of the components are analyzed.
Although not explicitly quoted in (\ref{lhd_sum}) we also include
into $L_{\Delta m^2}(r_1,r_2)$ the knowledge, that the delayed
energy $E_{\rm del}$ and the time difference $t_{\rm diff}$ differ for \CB\
(see Fig.~\ref{CC_n12b12}).

\begin{figure}
  \centerline{\epsfig{figure=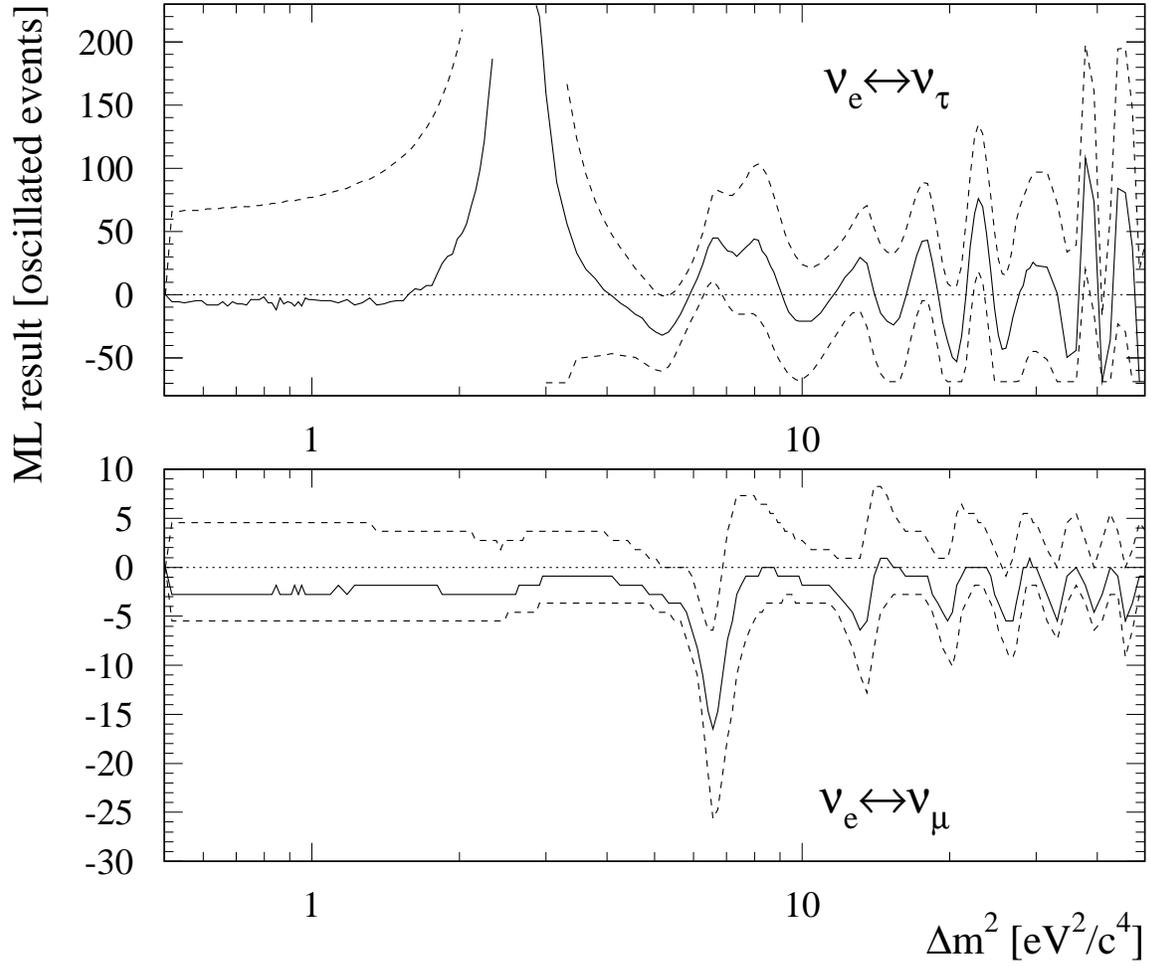,width=14.0cm}}
  \caption{Result of the 2--dimensional ML shape analysis of the 458 events
	in the channels
	\nuenutau\ and \nuenumu \,$+$\,\numubnueb . Shown are the 
	best fit values for oscillation contributions (solid lines) with the 
	corresponding 1--$\sigma$ errors (dashed lines) as a function of
	the oscillation parameter \Dm . }
  \label{CC_lhd}
  \end{figure}

Maximising $L_{\Delta m^2}(r_1,r_2)$ gives the most likely oscillation 
fractions $r_{1max}(\Dm)$, $r_{2max}(\Dm)$. This optimization is done 
separately for each value of \Dm\ in the range $0.1$ to 100\,\eVc . 
Over this range of \Dm\ the fully 2--dimensional best fit values are 
compatible with 
$r_1=r_2=0$ within the $1\sigma$ error band (see Fig.~\ref{CC_lhd}). 
As there is no indication for oscillations, upper limits
$r_1(\NCL)$, $r_2(\NCL)$  for each \Dm\ are determined.
In the case of \etau\ the ML analysis is not able to discriminate oscillation 
events for $2\lesssim \Dm \lesssim 3$\,\eVc\ and $\Dm>50$\,\eVc .
This is due to the fact that energy and spatial 
distributions for these oscillation cases are almost identical with the
expected electron spectra without oscillations. However, for \emu\ mixing, 
the shape analysis is also sensitive for large \Dm\ since we expect events
with $E_{pr}>36$\,MeV from \numubnueb .\footnote{The energy window 
$36<E_{pr}<50$\,MeV can in fact be used to search for \numubnueb\ oscillations
in the appearance mode with \nueb\ detection via \CB .} 
The \NCL\ limits for each \Dm\ are 
calculated by integrating the corresponding 2--dimensional likelihood 
function for $r_1$($r_2$) over the whole $r_2$($r_1$) range and then 
renormalizing to the physical region $r_1(r_2)>0$ applying a Bayesian approach 
near the physical boundary $r_1(r_2)=0$, respectively. The limits obtained 
are equivalent to the limits \Ptau(\NCL), \Pem(\NCL) in oscillation 
probability. The translation of these limits into \NCL\ limits for the 
neutrino oscillation amplitude and mixing angles for different values of
\Dm\ is shown in section~\ref{combi}.

  \section{Exclusion limits }\label{combi}

In this section, we translate the results of the different methods obtained 
in section~\ref{oscisearch} as limits on the
oscillation probability (cross section evaluation) or on 
the $\nu$--mixings (shape analysis) into limits of the oscillation amplitude. 
This oscillation amplitude $A$ is then expressed in terms of the mixing 
angles, for example \Aet\,=\,\sit\ (2 flavors) or 
$\Aet = 4sin^2\Phi cos^2\Phi cos^2\Psi$ (3 flavors, see section~\ref{intro}).

\subsection{Exclusion limits from measurement of the CC cross section}
	    \label{excl_CC}

We first discuss the limit obtained from the evaluation of the absolute 
number of detected CC sequences (section \ref{abscross}): 
$\Pex < 0.169$\,(\NCL)\footnote{Here, we do not consider the limit 
(\ref{CC_NC_result}) from section \ref{crossratio} 
since it is less stringent and therefore included in the limit used above.}
in a 2 flavor as well as in a complete 3 flavor scheme.

For a fixed neutrino energy and a fixed source--target distance $L$, the 
oscillation probability $P$ is related to the mixing amplitude $A$ via the
spatial evolution term in (\ref{osdef}). For an experimental 
configuration with continuous $\nu$ energies and a large volume detector with
its resolution functions, this spatial evolution term is determined by MC
simulations resulting in oscillation detection efficiencies
$\epsilon(\nu,\Dm)$ and $\epsilon(\bar{\nu},\Dm)$ as shown in Fig.~\ref{CC_eff}.
The upper limit of the oscillation amplitude for a given \Dm\ is then
  \begin{equation}
	\Aet \cdot \epsilon(\nu,\Dm) < \Ptau(\NCL)
  \label{aet_equ}
  \end{equation}
For \nuenumu\ the analogous MC simulations of positrons from \numubnueb\
based on $f^{\bar{\nu}}_{\Delta m^2}$ were performed resulting in 
$\epsilon(\bar{\nu},\Dm)$ so that
  \begin{equation}
   \Aem \cdot ( \epsilon(\nu,\Dm) - \epsilon(\bar{\nu},\Dm)\: ) < \Pem(\NCL)
  \label{aem_equ}
  \end{equation}
\Ptau(\NCL) and \Pem(\NCL) are the evaluated limits of the oscillation 
probability from the different analyses of section \ref{oscisearch}.

\begin{figure}
  \centerline{\epsfig{figure=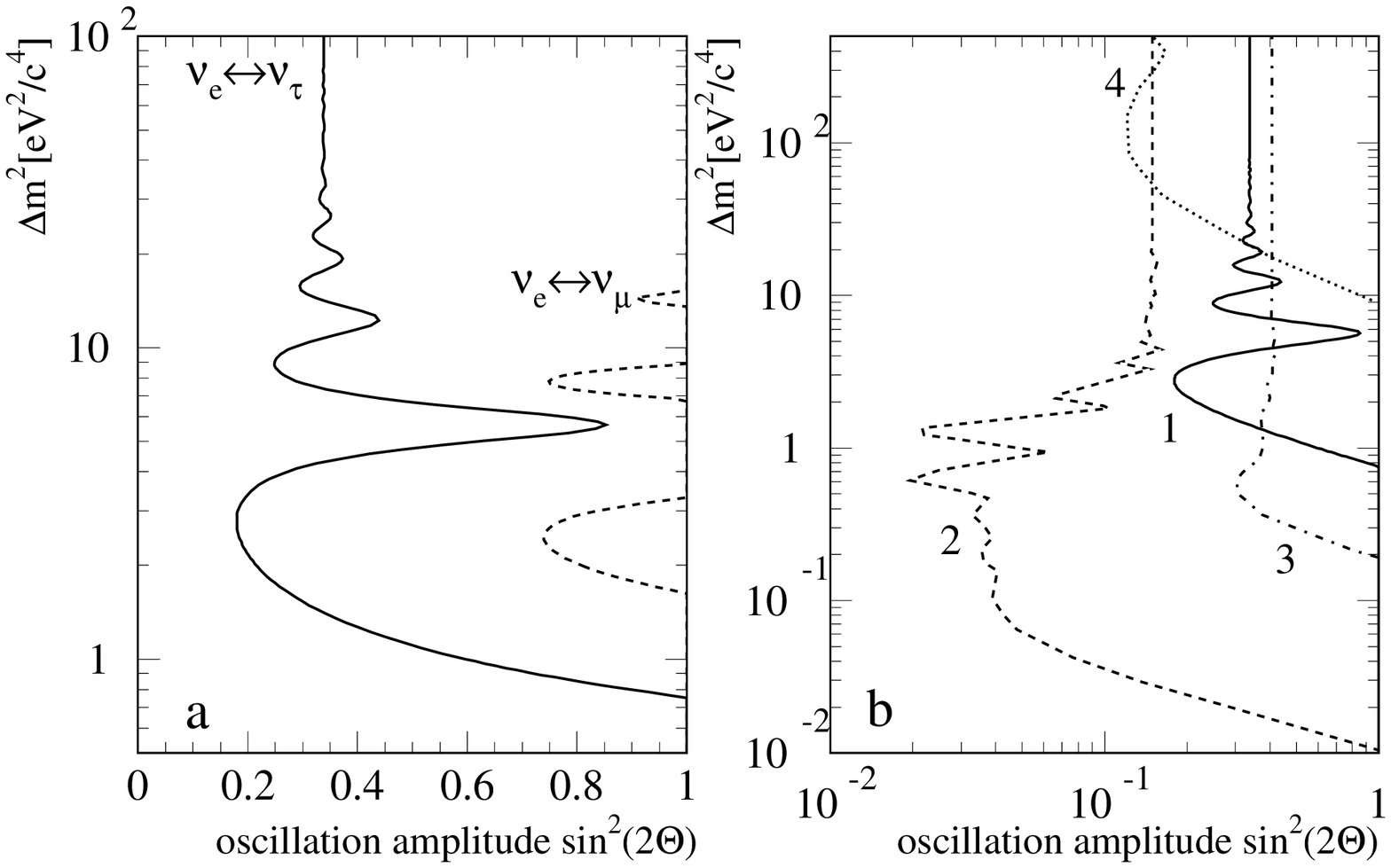,width=16.0cm}}
  \caption{2 flavor sensitivity plot for the different 
	channels \nuenumu\ and \nuenutau\ from KARMEN (a) and comparison with 
	limits of other experiments (b).\\
	a: KARMEN limits from the absolute cross section for \etau\ (solid 
	line) and \emu\ (dashed line) mixing.\\
	b: Limits from KARMEN \nuenutau(1), Bugey \nuebx(2), GALLEX \nuex(3) 
	and FNAL E531 \nuenutau(4). }
  \label{sensiplot}
  \end{figure}

Trivially, $\Pex = \Ptau + \Pem$ assuming three neutrino flavors.
However, in a 2 flavor scheme ignoring a second mixture, 
$\Pex = \Ptau$ or $\Pex = \Pem$
respectively. Therefore the \NCL\ limits in \sit\ can be evaluated following
(\ref{aet_equ}) and (\ref{aem_equ}): $\sit < 0.169/\epsilon(\nu,\Dm)$
for \nuenutau\ or $\sit < 0.169/( \epsilon(\nu,\Dm) - \epsilon(\bar{\nu},\Dm) )$ 
for \nuenumu . Figure \ref{sensiplot}a shows the exclusion curves from
this evaluation. Note that logical consistency requires that either one or
the other limit holds, not both simultaneously. 
Parameter combinations of \Dm\ and \sit\ to the right of 
the curves are excluded at \NCL . Taking \etau\ mixing, this corresponds to
mixing angles of $\sit>0.338$ excluded for large \Dm\ and $\Dm>0.77$\,\eVc\ 
excluded for maximal mixing. Due to the very small efficiency 
$\epsilon(\nu,\Dm) - \epsilon(\bar{\nu},\Dm)$ only three minor parameter regions
can be excluded assuming \emu\ mixing.
Figure \ref{sensiplot}b shows the KARMEN exclusion (curve 1) for 
\nuenutau\ together with exclusion limits from other experiments. Curve 2
represents the 90\% exclusion limit from a negative oscillation search in
the channel \mbox{\nueb $\rightarrow\,x$} at the nuclear power reactor of 
Bugey, France~\cite{bugey}. 
The second limit of a disappearance search comes from the 
GALLEX detector \cite{gallex} in which the \nue\ flux from the two
$^{51}$Cr source tests has been investigated\footnote{The complete analysis of
the two $^{51}$Cr source tests gives a $\nu$ signal of $R=0.93\pm 0.08$ compared
with the expectation and results in slightly higher upper limits than an
analysis \cite{bahcall} of a preliminary result of the first source test with 
$R=1.04\pm 0.12$.} with respect to \nuex .
The Fermilab experiment E531 searched for CC $\tau$--production through 
\nuenumu\ and \nuenutau\ appearance in a wideband \numu(\nue) beam~\cite{ush86}.
The exclusion limit from the \nuenutau\ search is shown as curve 4. 
Due to the small
experimental values $L/E <0.1$ with a most probable value of $L/E \approx 0.01$,
E531 is only sensitive to values of $\Dm > 10\eVc$.

\begin{figure}
  \centerline{\epsfig{figure=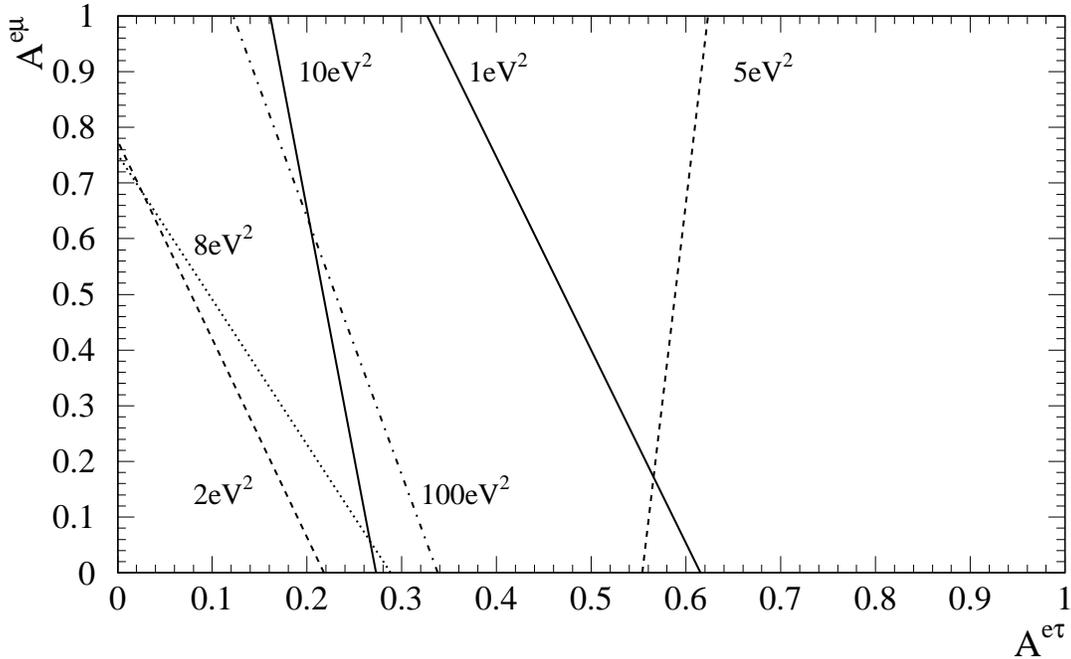,width=14.0cm}}
  \caption{3 flavor \NCL\ exclusion limits of the mixing amplitudes 
	$A^{e\mu}$ and $A^{e\tau}$ based on the extracted oscillation limit
	of \Pex $<$ 0.169 (\NCL) from the cross section analysis 
	(see equ.~\ref{CC_abs_result}).	Areas to the right of the 
	lines are excluded. }
  \label{CC_mix}
  \end{figure}

In a more detailed 3 flavor evaluation, the KARMEN oscillation limit \Pex\
corresponds to
  \begin{equation}
   \Aet \epsilon(\nu,\Dm) + \Aem ( \epsilon(\nu,\Dm) - 
	\epsilon(\bar{\nu},\Dm) ) = \Pex < \Pex (\NCL) = 0.169
  \label{abs_threedim}
  \end{equation}
In Fig.~\ref{CC_mix} the correlation of \Aet\ and \Aem\ limits is shown
for some examples of \Dm . The parameter \Aem\ is constrained
only for regions at \Dm\,$\approx$\,2--3\,\eVc , \Dm\,$\approx$\,7--9\,\eVc\
and at \Dm\,$\approx$\,13\,\eVc (see Fig.~\ref{sensiplot}a), whereas for 
\Dm\,$\approx$\,5\,\eVc\ the 
upper limit of \Aet\ even increases with \Aem\ due to the fact that 
$\epsilon(\nu,\Dm)-\epsilon(\bar{\nu},\Dm)$ is negative. 

\begin{figure}
  \centerline{\epsfig{figure=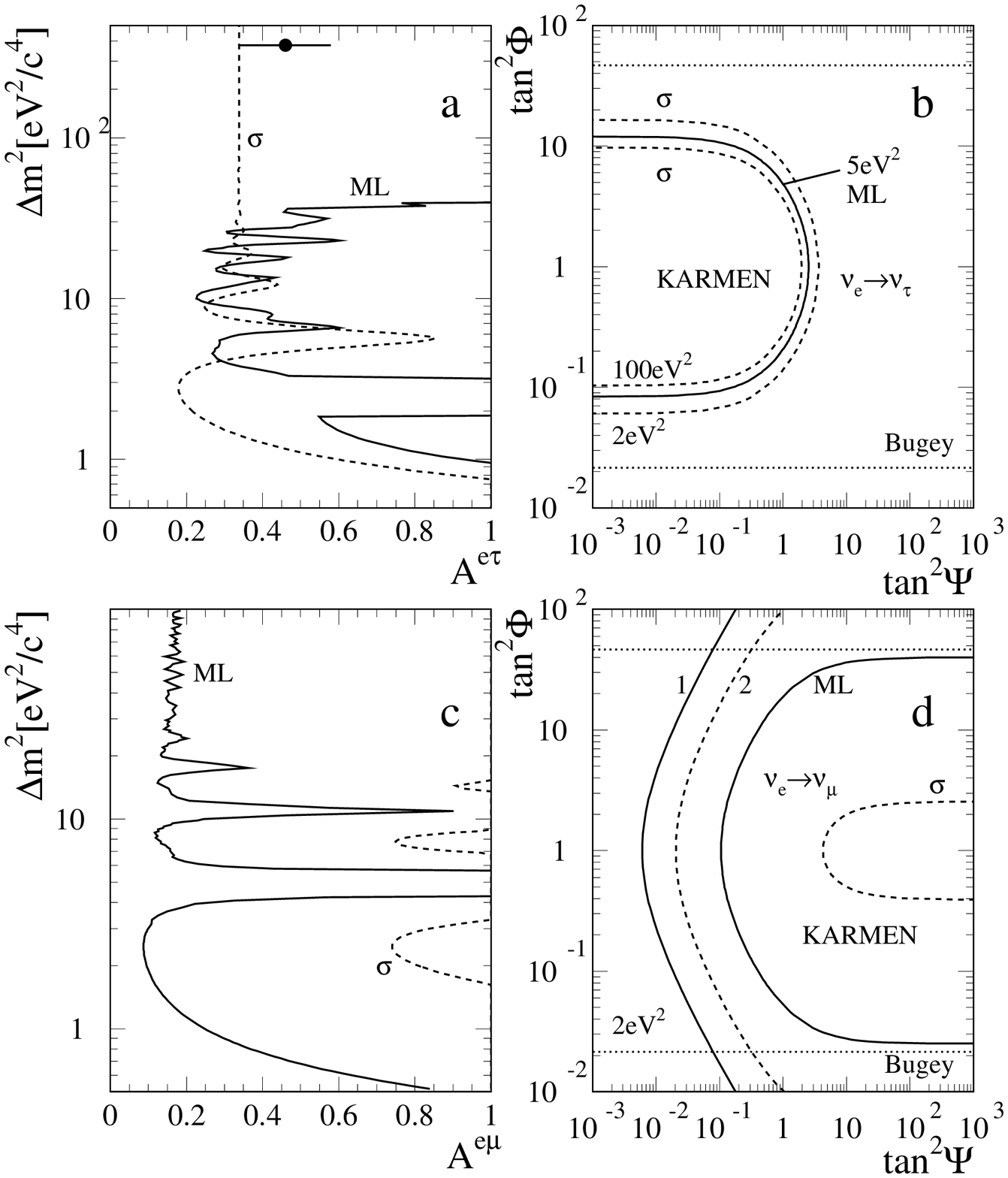,width=14cm}}
  \caption{a,c: KARMEN exclusion limits of the mixing amplitude $A^{e\tau}$ 
	and $A^{e\mu}$ (solid line = shape ML; dashed line ($\sigma$) = 
	absolute number of events)\\ b,d: 
	corresponding limits of the mixing angles $\Psi$ and $\Phi$ from the 
	limits of: (b) \Aet for several values of \Dm , (d) \Aem , for
	$\Dm=2$\,\eVc . For comparison, the exclusion limits from \nueb\ 
	disappearance search at Bugey \protect\cite{bugey} are shown, also for
	$\Dm=2$\,\eVc . In b(d) areas to the left(right) of the curves, 
	respectively, and between the horizontal lines are excluded. 
	In (d) curves 1 and 2 are the KARMEN limits from \numubnueb\ and
	\numunue\ appearance search. }
  \label{CC_s3detau}
  \end{figure}

From $\Aet = 4sin^2\Phi cos^2\Phi cos^2\Psi$ we can evaluate the \NCL\ limits 
for the mixing angles $\Phi$ and $\Psi$ taking the largest upper bound of \Aet\
with respect to \Aem . With the double logarithmic presentation of 
$tan^2\Phi,tan^2\Psi$ we follow again \cite{fogli}.
Figure~\ref{CC_s3detau}a,c shows the \NCL\ limits in \Aet\ and \Aem\ from the
cross section analysis as dashed curves (labeled with $\sigma$). 
In Fig.~\ref{CC_s3detau}b, these limits from \Aet\ are plotted as dashed 
curves for two examples of \Dm\ (2\,\eVc\ and 100\,\eVc) excluding areas in 
$(\Phi,\Psi)$ to the left. The corresponding limit on $(\Phi,\Psi)$ from \Aem\
for $\Dm =2$\,\eVc\ is shown in Fig.~\ref{CC_s3detau}d (dashed curve labeled
$\sigma$). 

The upper limit for \nuenutau , $\Ptau < 0.169$\,(\NCL), is reliable due to the
spectroscopic measurement of the \nue --flux from \mup\ DAR at ISIS and the
nearly background--free \excl\ detection reaction. This limit is in
contradiction to a model of 3 flavor neutrino mixing and masses calculated by
Conforto \cite{conf90,conf96}. Although being in conflict on a three standard
deviations level with the negative result in \nuenutau\ of FNAL E531
\cite{ush86} this compilation of different experimental results predicts
a transition probability $\Ptau=0.23\pm 0.06$\,(\NCL) expected in the 
appropriate 'short-baseline' regime. Figure~\ref{CC_s3detau}a
shows the calculated point $(\Dm ,\Aet)=(377\pm29\,\eVc ,0.46\pm0.12)$ lying
completely in the area excluded at \NCL\ by the KARMEN analysis of the 
absolute number of CC events.

\subsection{Exclusion limits from the spectral shape analysis}
	    \label{excl_shape}

In the 2--dimensional complete shape analysis (section~\ref{specshape})
the oscillation probabilities \Pem\ and \Ptau\ themselves depend on \Dm . 
The limit in the mixing amplitude is then 
  \begin{equation}
  \Aet < \frac{r_1(\NCL,\Dm)}{\epsilon(\nu,\Dm)} \qquad \mbox{and} \qquad
  \Aem < \frac{r_2(\NCL,\Dm)}{\epsilon(\nu,\Dm) - \epsilon(\bar{\nu},\Dm)}
  \end{equation}
The resulting limits in \Aet\ and \Aem\ are shown in Figure~\ref{CC_s3detau}a 
and Figure~\ref{CC_s3detau}c, respectively, as solid exclusion curves 
(labeled ML) together with the exclusion curves evaluated above (dashed lines).
For \nuenutau\ the analysis of the spectral shape of $L$ and $E$ is only 
sensitive and competitive to the evaluation of the absolute event number
in regions of about 0.5\,E/L $< \Dm <$ 5\,E/L corresponding to \Dm\ 
values of $\approx 3$ to 30\,\eVc\ for this experiment whereas for \emu\ the
shape analysis is far more sensitive. This is due to expected events from
\numubnueb\ with positron energies $E_{e^+}>36$\,MeV (see also 
Fig.~\ref{CC_n12b12}), where electrons from \excl\ cannot arise.
Fig.~\ref{CC_s3detau}b and Fig.~\ref{CC_s3detau}d show the regions in
$(\Phi,\Psi)$ space for various \Dm\ values and \Dm\,=2\,\eVc , respectively,
excluded by measurement of \etau\ and \emu\ mixing 
by KARMEN. These limits are compared with the exclusion limits
from the Bugey \nueb\ disappearance experiment where the particular value
of \Dm\ chosen for comparison, 2\,\eVc , has no special significance.
Note that there are no limits on \Aet\ from accelerator experiments like
FNAL E531 below \Dm\,=\,10\,\eVc .

\subsection{Global limits from the 3 flavor analysis }\label{excl_global}

\begin{figure}
  \centerline{\epsfig{figure=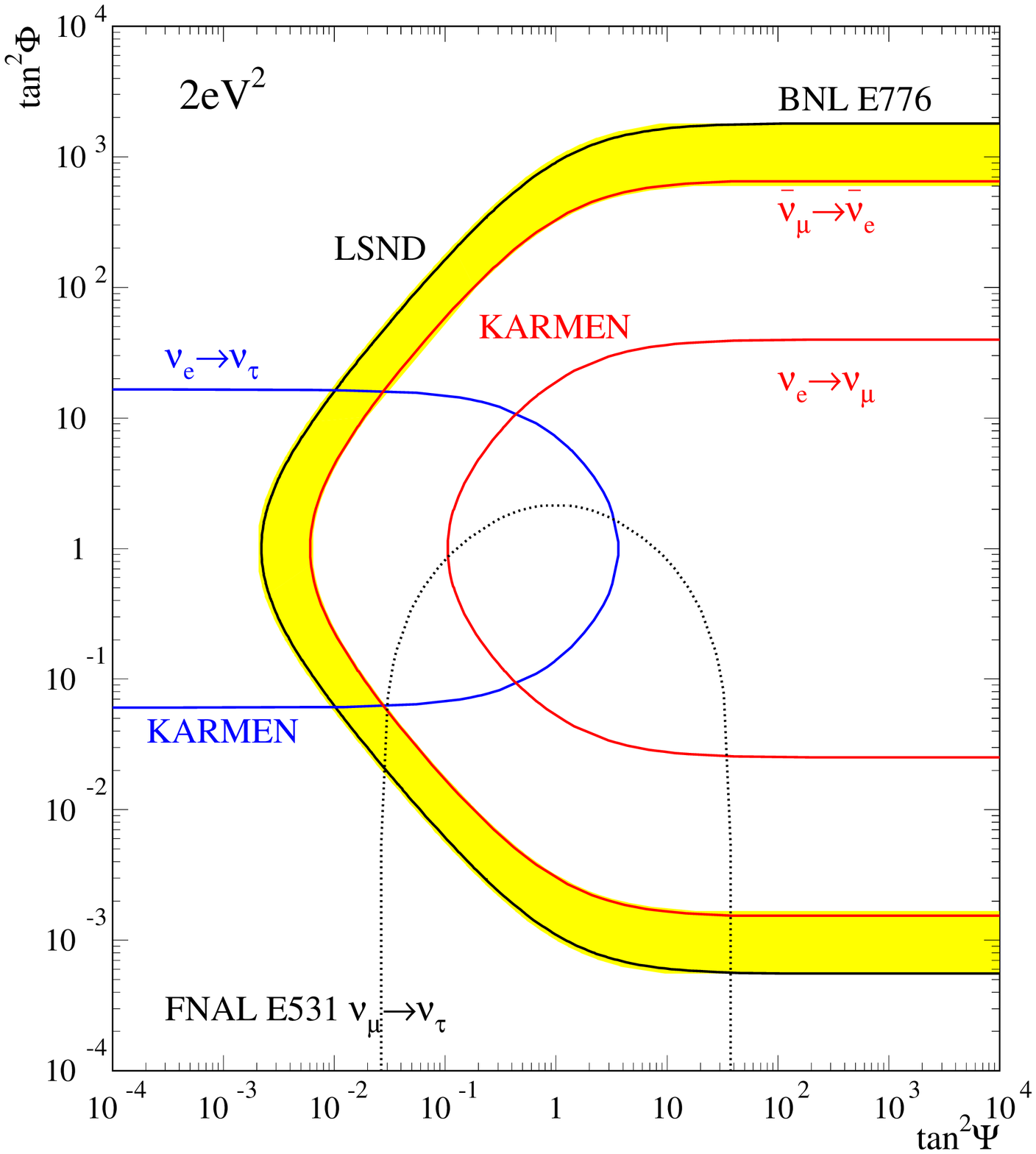,width=16.0cm}}
  \caption{\NCL\ limits to the mixing angles $\Phi$ and $\Psi$
	fixed at $\Dm = 2$\,\eVc\ deduced from this experiment from the
	oscillation channels \numubnueb\ appearance and \nuenumu , \nuenutau\ 
	disappearance
	in comparison with other experiments: FNAL E531 \numunutau\ appearance 
	\protect\cite{ush86}, BNL E776 \numubnueb\ 
	appearance \protect\cite{bnl776} and a 90\% likelihood favoured
	region from LSND \protect\cite{atha} (shaded region). }
  \label{allexp}
  \end{figure}

Up to now, there are no bounds from \nuenutau\ to the mixing angles $\Phi$, 
$\Psi$ in the region $\Dm \le 10$\,\eVc\ (see for example FNAL E531 limit in
Fig.~\ref{sensiplot}b).
With $\Aet =4U_{e3}^2 U_{\tau 3}^2 = 4 sin^2\Phi cos^2\Phi cos^2\Psi$ (see
equ.~\ref{osdef}) we can set new limits to the mixing angles $\Phi$ and $\Psi$ 
from \nuenutau\ oscillation search (see Fig.~\ref{CC_s3detau}b, \ref{allexp}). 
For a given \Dm\ of for example $\Dm = 2$\,\eVc\ the limits on the oscillation
amplitudes \Aet\ and \Aem\ from the \nuex\ search can be combined with other
limits from KARMEN as well as from other experiments.
Figure~\ref{allexp} demonstrates the complementarity
of different experimental results with respect to the mixing angles. 

The shaded area represents the 90\% likelihood regions obtained by LSND 
\cite{atha} in favour of \numubnueb\ oscillations. The other curves show \NCL\
limits from KARMEN (\numubnueb\ appearance search, \nuenumu\ and \nuenutau\
from this \nue\ disappearance search), BNL E776 (\numubnueb\ appearance 
\cite{bnl776}) and FNAL E531 (\numunutau\ appearance \cite{ush86}).
For this value of \Dm , the \numubnueb\ exclusion curve from BNL E776 just
covers the allowed region by LSND\footnote{Note that for values 
$\Dm < 2$\,\eVc\ BNL E776 with its \NCL\ limit does not cover the whole LSND 
region in ($\Phi$,$\Psi$) any more.} 
with the KARMEN \numubnueb\ sensitivity now substantially
improved following an experimental upgrade (see section \ref{conclu}). 
However, the complementary searches of KARMEN \nuenutau\
and FNAL E531 \numunutau\ clearly exclude regions of the LSND evidence with 
$tan^2\Phi \approx 1$ and $tan^2\Psi \approx 1$, respectively. 
Again, the necessity of investigating all oscillation channels to aim
for a consistent description of neutrino mixing becomes evident.

  \section{Conclusion and Outlook}\label{conclu}

The KARMEN experiment has found no positive evidence for $\nu$ oscillations in
the investigated channels \nuenutau\ and \nuenumu\ through disappearance of 
\nue . Charged current reactions induced by \nue\ were analyzed by three
different methods investigating the absolute number of events, the ratio of
CC to NC events and the spectral shape of energy $E$ and target distance $L$.
The sensitivity of the \nuex\ search is limited by the accuracy of the
theoretical cross section of \excl\ and the $\nu$ flux calculation in the
first evaluation, and by the relatively small statistics of 458 events in 
the case
of the shape analysis. The experimental parameter $L/E$ can be measured with
high accuracy on an event basis and differs significantly compared to other
\nueb\ disappearance searches at reactors or \nuenutau\ appearance searches
at accelerators.
In the maximum likelihood analysis, the 3 flavor approach assures fully
independent treatment of \etau\ and \emu\ mixing as free parameters. The
sensitivity of KARMEN is competitive with existing limits from
oscillation searches at reactors and improves limits from accelerator
experiments for \nuenutau\ for $\Dm < 10$\,\eVc .

In 1996, the KARMEN detector system was improved by an additional veto
counter system consisting of 300\,m$^2$ of plastic scintillator modules placed
within the massive iron shielding of the central detector. This new veto
system will reduce significantly cosmic induced background for the oscillation
search, especially in the appearance channel \numubnueb\ \cite{drex97}.
In the ongoing measuring period we expect again about 400 \nue --induced
CC sequences from \excl . Therefore, the statistical significance of this
\nuex\ investigation will slightly improve ending up with about 900
\nue --induced CC reactions in the KARMEN detector after a measuring period
of 2--3 years.

We gratefully acknowledge the financial support of the German 
Bundesministerium f\"ur Bildung, Wissenschaft, Forschung und Technologie 
(BMBF), the Particle and Astronomy Research Council (PPARC) and the Central 
Laboratory of the Research Council (CLRC) in the UK.

  \begin{table}
  \begin{center} \begin{tabular}{|l|c|c|c|l|}
model & \excl\ & \NC\ & R & reference \\ \hline \hline
EPT\footnotemark 
    & 9.19 & 9.87 & 0.93 & Fukugita~\cite{fuku}  \\ \hline
EPT & 9.0 & 10.6 & 0.85 & Mintz~\cite{mintz}  \\ \hline
CRPA\footnotemark
    & 9.29 & 10.53 & 0.88 & Kolbe~\cite{kolbe94}  \\ \hline
Shell with OBD\footnotemark & 9.1 & 9.8 & 0.93 & Engel~\cite{kolbe96}  \\ \hline
Shell with OBD\footnotemark & 9.4 & 9.9 & 0.95 & O'Connell~\cite{conn} \\ \hline
EPT & & 10.3 & & Bernab\'{e}u~\cite{berna}  \\ \hline
Shell & & 9.9 & & Parthasarathy~\cite{parth}  \\ \hline \hline
      & $9.4\pm 0.4\pm 0.8$ & $10.8\pm 0.9\pm 0.8$ & $0.86\pm 0.08$ 
      & this experiment \\ 
  \end{tabular} \end{center}
  \caption{Comparison of theoretical cross sections \msig{}{th} in units 
	of $10^{-42}$cm$^2$ averaged over the neutrino energy from \mup\ DAR 
	and the values \msig{}{exp} measured with KARMEN (including statistical 
	and systematical errors); for the NC cross section $\nu$=\nue +\numub\ 
	in case of no oscillations; R indicates the ratio \scc /\snc .}
  \label{crosstable}
  \end{table}
\addtocounter{footnote}{-3} \footnotetext{Elementary Particle Treatment} 
\stepcounter{footnote} \footnotetext{Continuum Random Phase Approximation} 
\stepcounter{footnote} \footnotetext{One Body Density}
\stepcounter{footnote} \footnotetext{This value for the CC exclusive cross 
section is calculated by \cite{kolbe96} using the OBD of \cite{conn}, but is 
also obtained with the original computer code NUEE from \cite{donn}.}


  \end{document}